\documentclass[prb,showpacs,floatfix,superscriptaddress,twocolumn]{revtex4-1}
\usepackage{amsmath}
\usepackage{graphicx}
\usepackage{float}
\usepackage{bm}
\usepackage{color}
\begin{document}

\bibliographystyle{apsrev}

\title{Optical Properties of Bogoliubov Quasiparticles}
\author{E. Schachinger}
\email{schachinger@itp.tu-graz.ac.at}
%\homepage{www.itp.tu-graz.ac.at/~ewald}
\affiliation{Institute of Theoretical and Computational Physics,
Graz University of Technology, A-8010 Graz, Austria}
\author{J. P. Carbotte}
\affiliation{Department of Physics and Astronomy, McMaster University, Hamilton,
Ontario, Canada N1G 2W1}
\affiliation{The Canadian Institute for Advanced Research, Toronto, Ontario,
  Canada M5G 1Z8}
\date{\today}
\begin{abstract}
We calculated the optical conductivity $\sigma(T,\Omega)$ of a gas of
Bogoliubov quasiparticles (BQP) from their Green's function and the
Kubo formula. We compare with corresponding normal state (N) and
superconducting state (SC) results. The superconducting case includes
the dynamic response of the condensate through additional contributions
to the Kubo formula involving the Gor'kov anomalous Green's function.
The differences in the optical scattering rate are largest just
above the optical gap and become progressively smaller as the photon
energy is increased or the temperature is raised. Our results are
compared with those obtained using a recently advocated phenomenological
procedure for eliminating the effect of the condensate.\cite{dord2014}
The $\delta$-function contribution at zero photon energy, proportional to
the superfluid density, is dropped in the real part of the conductivity
$[\sigma_1(T,\Omega)]$ and its Kramers-Kronig transform is subtracted from
the imaginary part $\sigma_2(T,\Omega)$. This results in deviations from
our BQP and superconducting state optical scattering rates even in the
region where these have merged and are, in addition, close to the normal
state result.
\end{abstract}
\pacs{74.25.nd, 74.20.-z, 74.25.fc}
\maketitle
%
% start paper
%

\section{Introduction}

Optical spectroscopy continues to find broad applicability as a technique
which reveals the dynamics of the charge carriers in a great variety of
materials. While other methods such as angular resolved photo emission
spectroscopy (ARPES) and scanning tunneling microscopy (STM) are
surface probes, optics involves the bulk. In superconducting materials the
real part of the complex dynamic longitudinal conductivity
$\sigma^{sc}(T,\Omega)$, at temperature $T$, contains combined information
on the absorption of photons of energy $\Omega$ from the Bogoliubov
quasiparticles (BQP) as well as from the condensate itself through
the braking of Cooper pairs. An important question raised recently is,
how can these two contributions be separated and, in particular, the
quasiparticle contribution $\sigma^{bqp}(T,\Omega)$ isolated? This
piece is directly related to quasiparticle relaxation processes which
are fundamental. Specifically, it speaks to the inelastic scattering
processes involved.

This work was motivated, in part, by a recent attempt by Dordevic
{\it et al.}\cite{dord2014} to separate out the quasiparticle contribution to the
optical conductivity [$\sigma^{sc}(T,\Omega)$] from that attributable
directly to the superfluid. The
phenomenological procedure advocated was to drop the Dirac delta-function at zero
photon energy (d.c.) which is proportional to the superfluid density $(\rho_s)$ and
enters in the real part of the optical conductivity. At the same time the
Kramers-Kronig transform of this contribution is to be subtracted from the
imaginary part of $\sigma^{sc}(T,\Omega)$. Their aim was to extract
directly from experimental optical conductivity data in the
superconducting state, information on quasiparticle relaxation processes.

Here we will base our considerations on theoretical results obtained from
Eliashberg theory which applies to an isotropic conventional electron-phonon
superconductor.
The dynamic longitudinal optical conductivity $\sigma^{sc}(T,\Omega)$ of a
superconductor as a function of temperature $(T)$ and photon energy $(\Omega)$
can be computed from the knowledge of the $2\times 2$
Nambu matrix Green's function\cite{nam1967a,nam1967b,mars2008} $(\cal G)$ and the
corresponding Kubo formula. The diagonal elements of the matrix $\cal G$
involve the quasiparticle Green's function or propagator $G$ with corresponding spectral
density $A({\bf k},\omega)$ where ${\bf k}$ is momentum and $\omega$ energy.
The Bogoliubov quasiparticles (BQP) are fully described by $A({\bf k},\omega)$,
which can be measured directly in angular resolved photo-emission experiments
(ARPES).\cite{damascelli2003} The question of the electrodynamics of the BQP gas
is, thus, well defined and follows from substituting $A({\bf k},\omega)$ into the
usual Kubo formula for $\sigma(T,\Omega)$ which we denote as $\sigma^{bqp}(T,\Omega)$.
The function $A({\bf k},\omega)$ also determines other properties of the BQP gas such as
the quasiparticle density of states, measured in scanning tunneling spectroscopy
(STM),\cite{fischer2007} and the electronic specific heat.\cite{mars2008}
In addition to terms involving the product of two spectral densities $A({\bf k},\omega)$,
the formula for the dynamic optical conductivity of the superconducting state further
contains products of two other spectral functions $B({\bf k},\omega)$. These relate to
the off diagonal elements of $\cal G$ and are the spectral functions for the Gor'kov
anomalous Green's functions $F$. The $F$ are related
to the long range order associated with Cooper pair formation. It is this second
contribution which makes $\sigma^{sc}(T,\Omega)$ different from $\sigma^{bqp}(T,\Omega)$.
In this paper we will be interested in computing separately the electrodynamics
of the BQP gas and in comparing it to the electrodynamics of the superconducting
state as well as of the underlying normal state.

One could consider many complications such
as the existence of anisotropy,\cite{leung1976,odon1995a,odon1995b} a
gap with $d$-wave symmetry,\cite{carb1995} non-phonon boson exchange
mechanisms like spin fluctuations\cite{basov2005,carb2011,allan2015} relevant
to the cuprates and iron-compounds, and energy dependence in the electronic
density of states.\cite{mitrovic1983,papa2015,flores2016} Other issues are
multiple band effects,\cite{mazin2009,paglione2010,basov2011,chub2015}
orbital fluctuations\cite{kontani2010} which have become prominent in the
iron superconductors, and nematicity\cite{vojta2009,lawler2010} to name only
a few. However, these are not essential elements in the present context and
it will be sufficient here to mainly consider a conventional, isotropic $s$-wave
electron-phonon superconductor which can be described by an isotropic
electron-phonon interaction spectral density $\alpha^2F(\omega)$.
Our aim is to compute in this simplified model the theoretical properties
of the BQP gas and to compare these with superconducting and normal state
results. Of particular interest will be the differences that exist between
these quantities and the circumstances under which they become comparable
in size and even merge. Certainly, at a few times the maximum phonon energy
plus the optical gap we would not expect large deviations between these
quantities to exist.

Most of the results we will present here are based on the
electron-phonon interaction spectral density $\alpha^2F(\omega)$ for
Pb.\cite{mars2008,mcmillan1965} This material is the prototype strong
coupling system with superconducting state properties that deviate
substantially from the universal values of weak coupling BCS theory.
To further test the general validity of our results we also considered
V$_3$Si an A15 compound which falls in the weak coupling limit and for
which the maximum phonon energy is $49\,$meV in contrast to the case of
Pb which is particularly soft with $\omega_{\rm max} = 11\,$meV.
Modifications arising from $d$-wave gap symmetry will also be briefly
illustrated in the specific case of the cuprate HgBa$_2$CuO$_{4+\delta}$
(Hg1201) based on the electron-boson interaction spectrum $I^2\chi(\omega)$
determined by Yang {\it et al.}\cite{yang2009}

In Sec.~\ref{sec:2} we present the formalism that is needed to calculate
the longitudinal dynamic optical conductivity $\sigma(T,\Omega)$ of an
electron-phonon system at any temperature $T$ and photon energy $\Omega$.
The formula involves solutions of the Eliashberg equations for the
renormalized frequencies $\tilde{\omega}(T,\omega)$ and the superconducting
gap function $\tilde{\Delta}(T,\omega)$. These equations are a set of
coupled integral equations for $\tilde{\omega}(T,\omega)$ and
$\tilde{\Delta}(T,\omega)$. The kernels in these equations which define
the material parameters that determine superconductivity, involve the
electron-phonon interaction spectral density $\alpha^2F(\omega)$ and
a Coulomb pseudo-potential parameter $\mu^\star$. For the convenience
of the reader these equations are discussed in Appendix \ref{app:B}.
They can be determined in the clean
limit and, later, an impurity residual scattering
rate $1/\tau_{\rm imp}$ will be introduced to deal with additional elastic
scattering. All inelastic scattering
is fully included in our clean limit solutions of the Eliashberg equations.
The BQP spectral density $A({\bf k},\omega)$ and the associated Gor'kov
$B({\bf k},\omega)$ are constructed from the renormalized frequencies
$\tilde{\omega}(T,\omega)$ and the gap function $\tilde{\Delta}(T,\omega)$.
The optical conductivity of the BQP gas $\sigma^{bqp}(T,\Omega)$ follows from
a knowledge of the $A({\bf k},\omega)$ alone. For $\sigma^{sc}(T,\Omega)$ the
Gor'kov $B({\bf k},\omega)$ also enter the Kubo formula. Normal state results
are obtained by setting the gap function $\tilde{\Delta}(T,\omega)\equiv 0$
everywhere.

In Sec.~\ref{sec:3} we present several relevant intermediate results which allow
us to calculate the optical scattering rate. This quantity which is related to
the imaginary part of the optical self-energy\cite{gotze1972} plays a dominant
role in the analysis of optical data. However, for the superconducting state
its precise interpretation in terms of quasiparticle relaxation remains
controversial.\cite{dord2014} We compare results for the superconducting state,
the BQP gas, the normal state, and with equivalent results obtained from the
prescription considered in the work of Dordevic {\it et al.}\cite{dord2014}
In Sec.~\ref{sec:4} we provide additional results for higher temperatures and
for the case when residual scattering is included. The specific case of
V$_3$Si is contrasted with Pb and the cuprate Hg1201 with a $d$-wave gap
symmetry is briefly discussed. Finally, Sec.~\ref{sec:5}
provides a discussion and conclusions.

\section{Formalism}
\label{sec:2}

In terms of the spectral density associated with the quasiparticle Green's
function $G$ (anomalous Gor'kov Green's function $F$) denoted by
$A({\bf k},\omega)$ [by $B({\bf k},\omega)$] the Kubo formula for the
real part of the dynamic optical conductivity in the superconducting state
at temperature $T$ and photon energy $\Omega$ reads:
\begin{widetext}
\begin{eqnarray}
  \textrm{Re}\{\sigma^{sc}(T,\Omega)\} &=& \frac{e^2}{2\Omega}\sum_{\bf k}
    v_F^2({\bf k})\int_{-\infty}^\infty\!\frac{d\omega}{2\pi}\left[
    f(T,\omega)-f(T,\omega+\Omega)\right]\nonumber\\
    &&\times\left[A({\bf k},\omega)A({\bf k},\omega+\Omega)+
      B({\bf k},\omega)B({\bf k},\omega+\Omega)\right],
\label{eq:1} 
\end{eqnarray}
\end{widetext}
with $e$ the charge on the electron, {\bf k} the momentum,
$v_F({\bf k})$ the Fermi velocity, and
$f(T,\omega)$ the Fermi-Dirac distribution function at temperature $T$.
The first term within the last set of square brackets,
$A({\bf k},\omega)A({\bf k},\omega+\Omega)$, defines the optical
conductivity of the BQP gas $[\sigma^{bqp}(T,\Omega)]$, while
the second term, $B({\bf k},\omega)B({\bf k},\omega+\Omega)$, is a
direct additional contribution to $\textrm{Re}\{\sigma^{sc}(T,\Omega)\}$
from the condensate and is not a part of the conductivity of a BQP gas.
The $A({\bf k},\omega)$ are determined by the renormalized frequencies
$\tilde{\omega}(T,\omega)$ and gap functions $\tilde{\Delta}(T,\omega)$
and, thus, know about the condensate. Consequently, the electrodynamics of
the BQP gas is an unambiguous concept and follows from retaining only the
first term in the last line of Eq.~\eqref{eq:1}.

After considerable algebra\cite{schur1998,mars2008} the superconducting
state complex optical conductivity of an electron-phonon metal
can be reduced to a convenient form for numerical calculations.
A particularly compact form given in the literature is:\cite{carb2005}
\begin{widetext}
\begin{equation}
  \label{eq:2}
  \sigma^{sc}(T,\Omega) = \frac{\Omega^2_p}{4\pi}\frac{i}{\Omega}
  \int_0^\infty\!d\omega\,\textrm{tanh}\left(\frac{\beta\omega}{2}
  \right)
  \left[J(\omega,\Omega)-J(-\omega,\Omega)\right],
\end{equation}
with $\Omega_p$ the plasma frequency, $\beta = 1/(k_BT)$, and
\begin{eqnarray}
  2J(\omega,\Omega) &=& \frac{1-N(\omega)N(\omega+\Omega)
  -P(\omega)P(\omega+\Omega)}{E(\omega)+
  E(\omega+\Omega)}\nonumber\\
  && +\frac{1+N^\star(\omega)N(\omega+\Omega)+
    P^\star(\omega)P(\omega+\Omega)}{E^\star(\omega)-
    E(\omega+\Omega)},
\label{eq:3}
\end{eqnarray}
\end{widetext}
where $\star$ indicates the complex conjugate. Equation~\eqref{eq:2}
gives both, the real and imaginary part of the dynamic conductivity. Here
\begin{eqnarray}
  E(\omega) &=& \sqrt{\tilde{\omega}^2(T,\omega+0^+)-
     \tilde{\Delta}^2(T,\omega+i0^+)},\nonumber\\
  E(-\omega) &=& -E^\star(\omega).
\label{eq:4}
\end{eqnarray}
By definition
\begin{equation}
   N(\omega) = \frac{\tilde{\omega}(T,\omega+i0^+)}
               {E(\omega)},
\label{eq:5}
\end{equation}
and
\begin{equation}
   P(\omega) = \frac{\tilde{\Delta}(T,\omega+i0^+)}
               {E(\omega)},
\label{eq:6}
\end{equation}
with $\tilde{\omega}(T,\omega+i0^+)$ and $\tilde{\Delta}(T,\omega+i0^+)$
the renormalized frequency and gap function determined from Eqs.~\eqref{eq:B1}
and \eqref{eq:B3}.
These follow in the usual way from the knowledge of the
electron-phonon interaction spectral density $\alpha^2F(\omega)$
associated with a particular material and the Coulomb repulsion parameter
$\mu^\star$. Here we will use the known spectrum of Pb in most of our
calculations.\cite{mars2008,mcmillan1965}
The conductivity of the BQP gas follows from the same
Eqs.~\eqref{eq:2} and \eqref{eq:3} but with all $P(\omega)$ terms left out.
%In the above
%$1/\tau_{\rm imp}$ is the static residual resistivity scattering rate
%and $\tilde{\omega}_{\rm clean}(\omega+i0^+)$ and
%$\tilde{\Delta}_{\rm clean}(\omega+i0^+)$ are the renormalized frequencies
%and the superconducting gap function of clean limit Eliashberg theory,
%respectively.

When considering the optical properties of superconductors it has become
common to introduce the concept of an optical self-energy or memory
function\cite{gotze1972}
\begin{equation}
  M(T,\Omega) = \Omega\left[\frac{m^\star_{\rm opt}(T,\Omega)}{m_b}-1\right]+
                \frac{i}{\tau_{\rm opt}},
\label{eq:7}
\end{equation}
with $\sigma(T,\Omega)$ related to $M(T,\Omega)$ such that
\begin{eqnarray}
  \frac{1}{\tau_{\rm opt}(T,\Omega)} &=& \frac{\Omega_p^2}{4\pi}\textrm{Re}\left\{
    \frac{1}{\sigma(T,\Omega)}\right\} \nonumber\\
    &\equiv&
    \frac{\Omega_p^2}{4\pi}\frac{\sigma_1(T,\Omega)}
    {\sigma_1^2(T,\Omega)+\sigma_2^2(T,\Omega)},
\label{eq:8}
\end{eqnarray}
and
\begin{eqnarray}
 \frac{m^\star_{\rm opt}(T,\Omega)}{m_b} &=& \frac{\Omega_p^2}{4\pi}\frac{1}{\Omega}
    \textrm{Im}\left\{\frac{1}{\sigma(T,\Omega)}\right\}\nonumber\\
    &\equiv&
    \frac{\Omega_p^2}{4\pi}\frac{1}{\Omega}\frac{\sigma_2(T,\Omega)}
    {\sigma_1^2(T,\Omega)+\sigma_2^2(T,\Omega)}.
\label{eq:9}
\end{eqnarray}
Here $\sigma_1(T,\Omega)$ and $\sigma_2(T,\Omega)$ are the real and imaginary
part of the complex optical conductivity $\sigma(T,\Omega)$, respectively,
and $m_b$ is the band mass. The scattering rate $1/\tau_{\rm opt}(T,\Omega)$ of
Eq.~\eqref{eq:8}, which is our primary interest in
this paper, is referred to as an {\em optical}
scattering rate while the effective mass $m^\star_{\rm opt}(T,\Omega)$ of Eq.~\eqref{eq:9},
which we will not consider, is an {\em optical} effective mass.
We will be interested only in finite photon
energies $\Omega+0^+$. Thus, we exclude here the Dirac delta-function which
appears at $\Omega=0$ in $\sigma_1^{sc}(T,\Omega)$ and which is proportional
to the superfluid density $\rho_s$. This conforms with the literature.

We consider four possible cases and compare these with each other, namely,
the superconducting state (sc), the Bogoliubov quasiparticle gas
(bqp), the normal state (n),
and results based on a phenomenological procedure aimed to extract
quasiparticle scattering rates from superconducting state optical
data (sc-m). The suggestion by Dordevic {\it et al.}\cite{dord2014} is
to take for the real part of the conductivity
\begin{equation}
  \sigma_1^{sc-m}(T,\Omega) = \sigma_1^{sc}(T,\Omega)-\rho_s\delta(\Omega),
\label{eq:10}
\end{equation}
and for the imaginary part their model conductivity
\begin{equation}
  \sigma_2^{sc-m}(T,\Omega) = \sigma_2^{sc}(T,\Omega)-\frac{\rho_s}{\Omega}.
\label{eq:11}
\end{equation}
These two equations are their equations (5) and (6) where
$\rho_s/\Omega$ is the Kramers-Kronig
transform of $\rho_s\delta(\Omega)$. We will denote the corresponding
optical scattering rate resulting from $\sigma^{sc-m}(T,\Omega)$ and
Eq.~\eqref{eq:8} by $\tau_{\rm opt}^{sc-m}(T,\Omega)$.

\section{Results at low temperature but with only inelastic scattering}
\label{sec:3}

\begin{figure}[t]
% \centering
\includegraphics[width = 95mm]{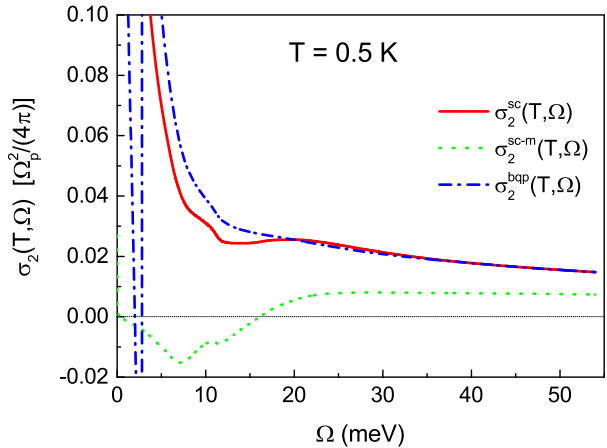}
\caption{(Color online) The imaginary part $\sigma_2(T,\Omega)$ of the optical
conductivity in units of $\Omega_p^2/(4\pi)$ as a function of the photon energy
$\Omega$. The temperature is $T = 0.5\,$K and the Pb electron-phonon spectral
density was used with a superconducting critical temperature $T_c=7.2\,$K. The
solid (red) curve is for the superconducting state (sc). The dotted (green)
curve presents the case when $\sigma_2(T,\omega)$ is modified according to the
prescription of Dordevic {\it et al.}\cite{dord2014} (sc-m). Finally, the
dashed-dotted (blue) curve presents the result for the Bogoliubov quasiparticles
(bqp).}
\label{fig:fig1}
\end{figure}

\begin{figure}[t]
 \centering
\includegraphics[width = 95mm]{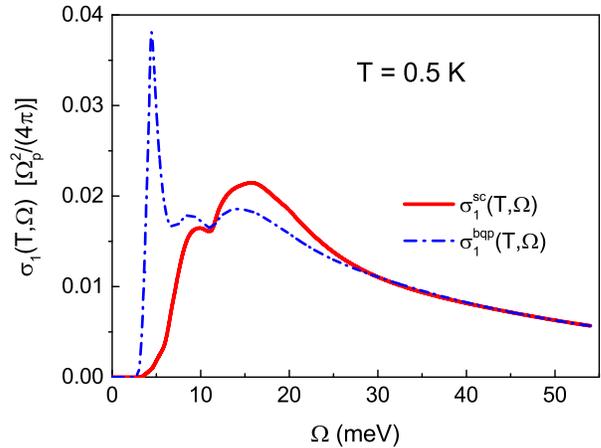}
\caption{(Color online) The real part $\sigma_1(T,\Omega)$ of the optical
conductivity in units of $\Omega_p^2/(4\pi)$ as a function of the photon energy
$\Omega$. The temperature is $T = 0.5\,$K and the Pb electron-phonon spectral
density was used. The solid (red) curve presents the result for the superconducting
state (sc) and the dashed-dotted (blue) curve is for the Bogoliubov quasiparticle
gas (bqp).}
\label{fig:fig2}
\end{figure}

In Fig.~\ref{fig:fig1} we present results for the imaginary part of
$\sigma(T,\Omega)$ in three cases. The solid (red) curve is the superconducting
case based on the electron-phonon interaction spectral function
$\alpha^2F(\omega)$ of Pb\cite{mcmillan1965} and our Eqs.~\eqref{eq:2}
and \eqref{eq:3} for the optical conductivity with the impurity scattering rate
$1/\tau_{\rm imp} = 0$. Thus, we are considering the clean limit and only inelastic
scattering is involved. The dashed-dotted (blue) curve is obtained from
Eqs.~\eqref{eq:2} and \eqref{eq:3} but including only BQP contributions,
i.e. $P(\Omega)$ is formally set equal to zero in Eq.~\eqref{eq:3} while all other
quantities remain unchanged. First, note that the main differences between these
two results are confined to the photon energies up to $\sim 20\,$meV or about
seven times the optical gap $2\Delta_0\simeq 2.8\,$meV.
Beyond this the two curves have merged. For the superconducting case,
$\sigma_2^{sc}(T,\Omega)$ goes to infinity as $\Omega$ tends to zero such that
the $\lim_{\Omega\to 0}\Omega\sigma_2^{sc}(T,\Omega) = \rho_s\Omega_p^2/(4\pi)$.
This limit provides a measure of the superfluid density $\rho_s$.
For the BQP gas and the model of Dordevic {\it et al.}\cite{dord2014}
there is, of course, no superfluid density.
It is very important for what will come later to note that for the
BQP gas $\sigma_2^{bqp}(T,\Omega)$ has a zero-crossing below
the optical gap. This does not occur in the superconducting
state. This zero will have serious manifestations when we raise
the temperature as we will in Sec.~\ref{sec:4}. The dotted
(green) curve gives $\sigma_2^{sc-m}(T,\Omega)$ obtained from Eq.~\eqref{eq:11}
according to the suggestion of Dordevic {\it et al.}\cite{dord2014} We see
that it differs from the other two curves at all photon energies and,
in particular, does not merge with the solid (red) and dashed-dotted
(blue) curves even at the highest energy considered here
($\sim 50\,$meV), where it is smaller by more than a factor of two.
This has important implications for the behavior of the optical scattering rate.
To understand why this is so we need to also consider the real part of
the optical conductivity for these three cases.
This is shown in Fig.~\ref{fig:fig2}.

\begin{figure}[t]
 \centering
\includegraphics[width = 95mm]{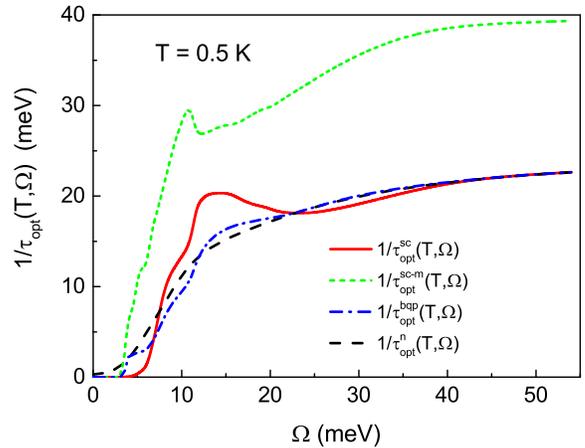}
\caption{(Color online) The optical scattering rate $1/\tau_{\rm opt}(T,\Omega)$
which is related to the optical self-energy as a function of the photon energy
$\Omega$. The temperature is $T = 0.5\,$K and the Pb electron-phonon spectral
density was used. The solid (red) curve is the superconducting state result (sc).
The dotted (green) is for the superconducting state modified according to the
prescription of Dordevic {\it et al.}\cite{dord2014} (sc-m). The dashed-dotted
(blue) curve presents the result for the Bogoliubov quasiparticle gas (bqp) and
the dashed (black) curve applies to the normal state (n).}
\label{fig:fig3}
\end{figure}

Only two curves are considered because $\sigma_1(T,\Omega)$ is the same for the
sc and sc-m cases. The solid (red) curve applies to the superconducting
state while the dashed-dotted (blue) curve is for the BQP gas. Optical absorption
starts at the finite photon energy $\Omega = 2\Delta_0\simeq 2.8\,$meV because
in both cases the quasiparticle excitation spectrum is gaped.
Just above the optical gap the two responses show distinct behavior.
The dashed-dotted (blue) curve for BQP has an inverse square root singularity
associated with the onset of optical absorption while the solid (red) curve
for the superconducting state, by contrast, rises slowly out of zero. In
Appendix~\ref{app:A} we calculate the real part of the optical conductivity of a BQP
gas in the clean limit of BCS theory. In this limit the spectral functions
in Eq.~\eqref{eq:1} are Dirac delta-functions and the integrals can be done
analytically. For the BQP case only the $A({\bf k},\omega)A({\bf k},\omega+\Omega)$
term appears. It results in two distinct
contributions. An intraband and an interband piece. The intraband part does not
contribute at $T=0$ as there is zero optical spectral weight in this
Drude-like term in the clean limit. The interband contribution has the
form~\eqref{eq:a6} (see Appendix~\ref{app:A})
\begin{equation}
 \label{eq:13}
 \textrm{Re}\{\sigma^{bqp}_{\rm inter}(T=0,\Omega)\} \propto
 \frac{\Lambda}{\Omega^2}\frac{\Delta^2}{\sqrt{\Omega^2-(2\Delta^2)}},
 \quad\textrm{for}\quad \Omega>2\Delta,
\end{equation}
which we repeat here for the convenience of the reader. This inverse square
root singularity is clearly seen in the dashed-dotted (blue) curve of
Fig.~\ref{fig:fig2}. As we included inelastic scattering
in our calculations, Eq.~\eqref{eq:13} is superimposed on a
phonon assisted background. This background extends to large energies while the
clean limit type contribution~\eqref{eq:13} is confined mainly to the
region $\Omega=2\Delta$ to $4\Delta$.
In contrast, the solid (red) curve for the superconducting
state shows no interband divergence. This is traced to the fact that
when the $A({\bf k},\omega)A({\bf k},\omega+\Omega)$ and
$B({\bf k},\omega)B({\bf k},\omega+\Omega)$ in Eq.~\eqref{eq:1} are
both included, the interband pieces cancel. (Class II matrix elements.)
Consequently, only the phonon assisted processes are seen in the solid
(red) curve. There is a phonon kink (a dip) in both (bqp) and (sc) curves
at an energy which corresponds to the end of the phonon spectrum plus the
optical gap. which falls around $\sim 13\,$meV. In the region beyond this
energy the two curves of Fig.~\ref{fig:fig2} show some deviations (never
large) from each other but at $\sim 25\,$meV, or $\sim 8$ times the
optical gap, they merge. It is important to understand that the
interband transitions involved in Eq.~\eqref{eq:13} have
nothing to do with relaxation processes but rather have mainly to do with
the electronic band structure of the BQP gas, and exist even when no scattering
is accounted for. Of course, the  high energy part of the curves reflects
the inelastic scattering and this region does carry information on
relaxation processes which is our primary interest in this work.
In the paper by Dordevic {\it et al.}\cite{dord2014}
the real part of the quasiparticle conductivity defined in Eq.~\eqref{eq:10}
is that given by the solid (red) curve of Fig.~\ref{fig:fig2}. Thus, in all three
cases of interest $\sigma_1(T,\Omega)$ is the same at high energies.
($\Omega\ge 25\,$meV). However, as we
have seen in Fig.~\ref{fig:fig1} only the superconducting state and BQP case merge
when we consider the imaginary part of the optical conductivity
$\sigma_2(T,\Omega)$ while the modified case
of Dordevic {\it et al.}\cite{dord2014} strongly deviates. This implies
similar differences for the optical scattering rate defined in Eq.~\eqref{eq:8} as
is seen in Fig.~\ref{fig:fig3}.
Staying for now with the high energy region we note, as expected, that the
solid (red) and the dashed-dotted (blue) curves match perfectly
while the dotted (green) curve for the modified superconducting case is larger
than the other two by a factor of two.
This can be understood entirely from the fact that
$\sigma_2^{sc-m}(T,\Omega)$ is much smaller in this energy range
than is $\sigma_2^{sc}(T,\Omega)$ or $\sigma_2^{bqp}(T,\Omega)$.
We included in Fig.~\ref{fig:fig3}, for reference, normal state
results as the dashed (black) curve. It agrees well with the superconducting
state and the BQP gas but not with the dotted (green) curve which we will,
therefore, not discuss further here because it does not
conform to our expectation that at large $\Omega$ the presence
of a gap in the quasiparticle excitation spectrum should no longer matter.

The deviations between the remaining three curves are in fact rather small
at all energies except for two features. First, the solid (red) curve for
the superconducting state shows a peak in the phonon assisted region
between $\sim 10\,$meV to $\sim 20\,$meV which is not part of the other
two curves. The deviations are of the order of $20\,\%$.
Second, at energies below $\sim 6\,$meV the solid (red) curve rapidly goes
to zero while the dashed-dotted (blue) curve shows a plateau before it
eventually drops to zero. This plateau has its origin in the interband transitions.
These are shown in Fig.~\ref{fig:fig2} as a large prominent peak between, roughly,
the energies $2\Delta(T)$ and $4\Delta(T)$. Returning to Fig.~\ref{fig:fig3},
below the optical gap the normal state dashed (black) curve is, of course, non-zero
as there is no gap in this case.
\begin{figure}[t]
 \centering
\includegraphics[width = 95mm]{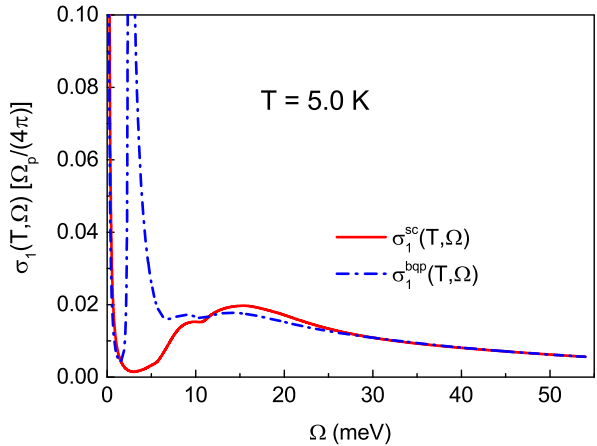}
\caption{(Color online) Same as Fig.~\ref{fig:fig2} but now the
temperature was raised to $T=5\,$K. The Pb electron-phonon spectral
density was used. The solid (red) curve applies to the superconducting state (sc)
and the dashed-dotted (blue) curve to the Bogoliubov quasiparticle gas (bqp).}
\label{fig:fig4}
\end{figure}

\section{Finite temperature and impurity scattering}
\label{sec:4}

In this section the temperature was chosen to be large enough
to expect results different from the $T=0.5\,$K case which
is basically the zero temperature limit. In particular, we chose $T=5\,$K
or $T/T_c \simeq 0.69$. This remains within the superconducting state but
is sufficiently high that the optical gap is reduced from its $T=0$ value
of $2.8\,$meV to $2.4\,$meV. Results for the real part of the optical
conductivity are presented in Fig.~\ref{fig:fig4}. The dashed-dotted (blue)
curve is for the BQP gas and the solid (red) curve is for the superconducting
state. The change in the superconducting gap is most clearly seen in the
dashed-dotted (blue) curve when compared with the equivalent curve of
Fig.~\ref{fig:fig2}. There is now finite absorption at all $\Omega$ below the gap
due to the existence of a small number of thermally excited quasiparticles
at this elevated temperature, in contrast to the $T=0$ case for which
the real part (absorptive part) of the optical conductivity is zero below
the optical gap. The phonon region above the optical gap is, in contrast, not
changed much from $T=0.5\,$K.

\begin{figure}[t]
\includegraphics[width = 95mm]{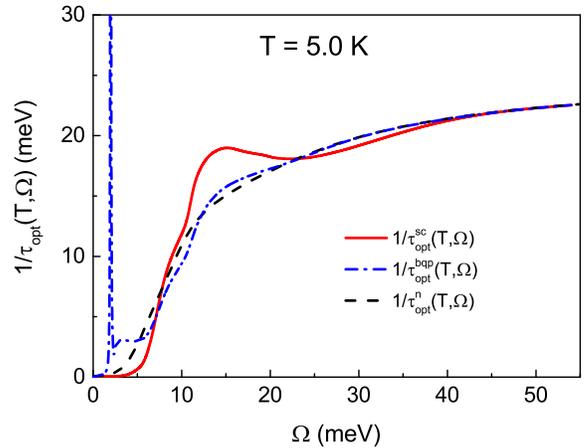}
\caption{(Color online) The same as Fig.~\ref{fig:fig3} for the optical scattering
rate $1/\tau_{\rm opt}(T,\Omega)$ but now for $T=5\,$K. The solid (red) curve applies
to the superconducting state (sc), the dashed-dotted (blue) curve to the Bogoliubov
quasiparticle gas (bqp), and the dashed (black) curve to the normal state (n).}
\label{fig:fig5}
\end{figure}

Results for the optical scattering rate are presented in Fig.~\ref{fig:fig5}.
The three curves are $1/\tau^{sc}_{\rm opt}(T,\Omega)$, superconducting
state [solid (red) curve], $1/\tau^{bqp}_{\rm opt}(T,\Omega)$, BQP gas
[dashed-dotted (blue) curve], and $1/\tau^{n}_{\rm opt}(T,\Omega)$, normal
state [dashed (black) curve]. These results are not very different
from those of Fig.~\ref{fig:fig3}, normal state and BQP gas remain close
above $6\,$meV. The superconducting state retains its peak in the range
of 10 to $20\,$meV and the plateau just above the optical gap in the
dashed-dotted (blue) curve remains as well. There is, however, a striking difference.
A sharp peak has developed below the optical gap at $\Omega_0\simeq 2\,$meV.
This is exactly the energy at which $\sigma_2^{bqp}(T,\Omega_0)$ crosses
zero. But, as seen in Fig.~\ref{fig:fig4}, $\sigma_1^{bqp}(T,\Omega_0)$ is
non-zero at this crossing and this is the origin of the peak seen in
$\tau_{\rm opt}^{bqp}(T,\Omega_0)$. As we have shown, $\sigma_2^{bqp}(T,\Omega)$
also has a zero-crossing (Fig.~\ref{fig:fig1}) at $T=0.5\,$K. This zero-crossing
has no measurable consequences in the optical scattering rate however,
because $\sigma_1^{bqp}(T\approx0,\Omega)$ is zero below the optical gap in
this case.

\begin{figure}[t]
 \centering
\includegraphics[width = 95mm]{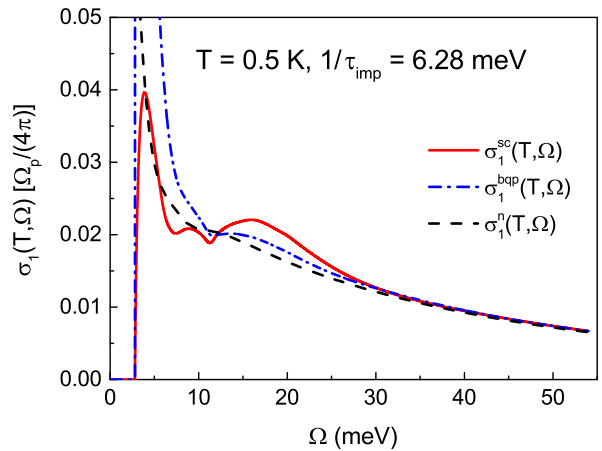}
\caption{(Color online) The same as Fig.~\ref{fig:fig2} for the real part of the
optical conductivity, But now elastic impurity scattering with
$1/\tau_{\rm imp} = 6.28\,$meV is included. The scale on the vertical axis has
been adjusted to make clear the Drude-like contribution seen (as a peak) just
above the optical gap. This peak is not there when $1/\tau_{\rm imp}=0$, i.e.
in the clean limit of inelastic scattering only. The solid (red) curve
applies to the superconducting state (sc), the dashed-dotted (blue) curve to the
Bogoliubov quasiparticle gas (bqp), and the dashed (black) curve to the normal
state (n).}
\label{fig:fig6}
\end{figure}

So far we did not include in our work static elastic residual impurity
scattering, rather only inelastic scattering was considered.
Eqs.~\eqref{eq:2} and \eqref{eq:3} for the optical conductivity were written
with a $i/\tau_{\rm imp}$ term and we now set  $1/\tau_{\rm imp} = 6.28\,$meV
in Eqs.~\eqref{eq:B3} for illustrative purposes.
This affects both the real and imaginary part of the optical
conductivity. In Fig.~\ref{fig:fig6} we present results for the real part of the
optical conductivity in the superconducting state. The solid (red)
curve has now acquired a sharp peak just above the optical gap which was not
there in Fig.~\ref{fig:fig2} for the pure case ($1/\tau_{\rm imp} = 0$). This
peak is what remains when the Drude peak of the normal state shown
as the the dashed (black) curve is cut off by the optical gap. In BCS theory this
would be the only contribution to the optical conductivity. Here, the inelastic
scattering provides additional absorption at higher energies. When
$1/\tau_{\rm imp}=0$ this is the only contribution that remains.
The Drude peak about $\Omega=0$ comes from the coherent part of
the quasiparticle Green's function $G$. At zero temperature and with no
elastic scattering this contribution collapsed into a Dirac delta-function at
the origin and, consequently, is not seen in Fig.~\ref{fig:fig2}.
It contains approximately $1/(1+\lambda)$ of the optical spectral weight,
where $\lambda$ is the electron-phonon mass enhancement parameter. This
contribution broadens out into a Drude peak when $1/\tau_{\rm imp}$ is non-zero.
To a good approximation this part of the normal state optical response can be
written at $T=0$ as\cite{mars2008}
\begin{equation}
  \sigma_1^{\rm Drude}(T,\Omega) = \frac{\Omega_p^2}{4\pi}
    \frac{\Gamma/(1+\lambda)}{\Omega^2+[\Gamma/(1+\lambda)]^2}
    \frac{1}{1+\lambda},
\label{eq:12}
\end{equation}
with $\Gamma \equiv 1/\tau_{\rm imp}$. In the superconducting state
this gets gaped with loss optical spectral weight transferred to the condensate.
The phonon assisted processes (incoherent part)
provide an additional contribution which contains the remaining optical spectral weight
approximately proportional to $\lambda/(1+\lambda)$ and these are
the only processes that show up at finite photon energies in
Fig.~\ref{fig:fig2} (pure case). Comparing the real part of the
optical conductivity for the BQP gas $\sigma_1^{bqp}(T\approx0,\Omega)$
[dashed-dotted (blue) curve] with impurities to its value in the clean
limit shown in Fig.~\ref{fig:fig2}, we note that the interband peak just
above the optical gap has been broadened. Also the secondary peak just
below $10\,$meV in Fig.~\ref{fig:fig2} has been modified to a shoulder.

\begin{figure}[t]
\includegraphics[width = 95mm]{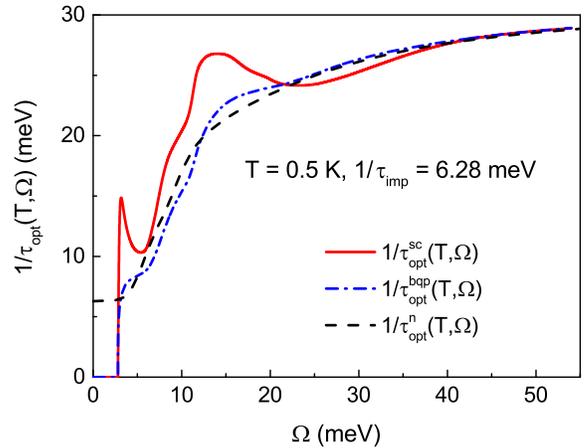}
\caption{(Color online) The same as Fig.~\ref{fig:fig3} for the optical scattering
rate $1/\tau_{\rm opt}(T,\omega)$ including elastic impurity scattering
$1/\tau_{\rm imp} = 6.28\,$meV. The solid (red) curve applies to the superconducting
state (sc), the dashed-dotted (blue) curve to the Bogoliubov quasiparticle gas
(bqp), and the dashed (black) curve to the normal state (n).}
\label{fig:fig7}
\end{figure}

The truncated Drude peak in the solid (red) curve of Fig.~\ref{fig:fig6}
also shows up as a peak in the optical scattering rate of Fig.~\ref{fig:fig7}
immediately above the optical gap.
The normal state $1/\tau_{\rm opt}^n(T,\Omega)$ [dashed (black) curve], 
however, has no optical gap and, consequently, no peak. However, it is
finite at $\Omega=0$ and equal to the residual scattering rate. In
general, it is the sum of $1/\tau_{\rm imp}$ plus the inelastic contribution of
the previous section (pure case). At low photon energies the curve is flat and equal
to $1/\tau_{\rm imp}$ because the inelastic contributions start from zero
at $\Omega=0$ and remain small.
In the other two curves we see that the absorption edge at
$\Omega=2\Delta(T)$
has effectively sharpened, as compared with the pure case of Fig.~\ref{fig:fig3},
showing an initial vertical rise out of zero. At high
energies the solid (red) curve, the dashed-dotted (blue) curve, and the
dashed (black) curve merge. BQP and normal state curves are in fact close to
each other at all photon energies above the optical gap. As before the superconducting
case shows a peak in the phonon assisted region which is not part of the other
two cases.

\begin{figure}[t]
\includegraphics[width = 95mm]{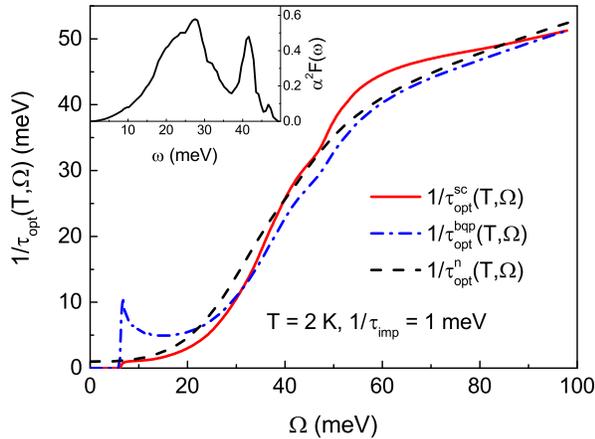}
\caption{(Color online) The same as Fig.~\ref{fig:fig7} (which was for Pb)
but now the V$_3$Si electron-phonon interaction spectral density was used
with an impurity scattering of $1/\tau_{\rm imp} = 1\,$meV. The critical
temperature $T_c=17.1\,$K, the temperature $T=2\,$K, and the gap
$\Delta_0 = 2.9\,$meV.  The solid (red) curve applies to the superconducting
state (sc), the dashed-dotted (blue) curve to the Bogoliubov quasiparticle gas
(bqp), and the dashed (black) curve to the normal state (n). The inset gives
$\alpha^2F(\omega)$ for V$_3$Si.\cite{carb1990} The essential difference
with Pb is that here the spectrum extends to $49\,$meV rather than
$\sim 11\,$meV.}
\label{fig:fig7a}
\end{figure}

In Fig.~\ref{fig:fig7a} we present additional results based on the
electron-phonon interaction spectral density if V$_3$Si.\cite{carb1990}
(See inset.)
In contrast to Pb, this material is close to the weak coupling limit and
also has a large maximum phonon energy of $\omega_{\rm max} = 49\,$meV.
The superconducting gap at $2\,$K is $\Delta_0 = 2.9\,$meV. The impurity
scattering was taken to be $1/\tau_{\rm imp} = 1\,$meV which remains in
the clean limit, i.e. $\Delta_0\tau_{\rm imp} > 1$ to be compared with
the results presented in Fig.~\ref{fig:fig7} for Pb where the dirty limit
applies, i.e. $\Delta_0\tau_{\rm imp} \ll 1$ with $\Delta_0 = 1.4\,$meV
and $1/\tau_{\rm imp} = 6.3\,$meV. In the clean limit of Fig.~\ref{fig:fig7a}
the rise in $1/\tau^{sc}(T,\Omega)$ [solid (red) curve] at the optical gap
is small as compared to that in Fig.~\ref{fig:fig7} because most of the
Drude peak is now below the gap. For the BQP case [dashed-dotted (blue)
curve] there is a
sharp rise and a peak followed by a plateau which extends to approximately
$20\,$meV before it starts to rise and crosses below $1/\tau^{sc}(T,\Omega)$
around $35\,$meV after which the two curves are never far apart. Except
for a scale difference related to $\omega_{\rm max}$ these results are
very similar to those of Pb. The physics is the same for weak or strong
coupling and small or large  maximum phonon energies $\omega_{\rm max}$.

\begin{figure}[t]
\includegraphics[width = 95mm]{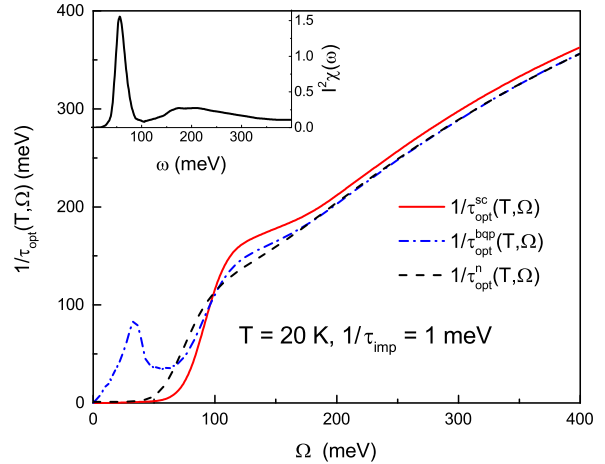}
\caption{(Color online) The same as Fig.~\ref{fig:fig7a} but now the
electron-boson interaction spectral density $I^2\chi(\omega)$ (see inset)
obtained by Yang {\it et al.}\cite{yang2009} for Hg1201 is used and the
gap is taken to have $d_{x^2-y^2}$ symmetry which is apptopriate for the
cuprates. The $T_c = 90\,$K, the temperature $T=20\,$K, the elastic
impurity scattering is $1/tau_{\rm imp} = 1\,$meV, and the gap is
$\Delta_0 = 19.5\,$meV. The solid (red) curve applies to the superconducting
state (sc), the dashed-dotted (blue) curve to the Bogoliubov quasiparticle
gas (bqp), and the dashed (black) curve to the normal state (n).
}
\label{fig:fig7b}
\end{figure}

In Fig.~\ref{fig:fig7b} we present a last set of results that illustrate
the effect of $d$-wave gap symmetry on the BQP optical scatterin rate.
The calculations are based on the electron-boson interaction spectral
density $I^2\chi(\omega)$ (see inset) obtained in the work by Yang
{\it et al.} for Hg1201.\cite{yang2009} To arrive at these results
the Eliashberg equations of Appendix \ref{app:B} had to be generalized
to include gap anisotropy. These are found in Ref.~\onlinecite{schach2003}
along with the appropriate generalizations of Eqs.~\eqref{eq:2} and
\eqref{eq:3}. Comparison of these results with those of Fig.~\ref{fig:fig7a}
reveals the same qualitative behaviour except that the gap is distributed
between $\Omega=0$ to $\Omega = 2\Delta(T) = 19.5\,$meV in accordance
to $d_{x^2-y^2}$ gap symmetry. The plateau above the gap remains and
beyond $\Omega\simeq 90\,$meV $1/\tau^{bqp}_{\rm opt}(T,\Omega)$ becomes
close to $1/\tau^{sc}_{\rm opt}(T,\Omega)$ and $1/\tau^{n}_{\rm opt}(T,\Omega)$.
The results presented are based on the optical conductivity within
the CuO plane ($ab$-plane).

\section{Conclusions}
\label{sec:5}

We calculated the optical conductivity $\sigma^{sc}(T,\Omega)$
(superconducting state) and $\sigma^{bqp}(T,\Omega)$ (BQP gas) and
compared the two. Of particular interest in such a comparison was the derived
optical scattering rate which corresponds to the imaginary part of the
optical self-energy. This optical constant was often highlighted in the
experimental literature because it provided information on quasiparticle
relaxation. What was directly accessible in experiment was the superconducting
state optical scattering rate $1/\tau^{sc}_{\rm opt}(T,\Omega)$ while it
was the BQP gas scattering rate $1/\tau^{bqp}_{\rm opt}(T,\Omega)$ which
was more directly interpretable. Thus, the question arose as to how similar or
different these two rates were. This was recognized in a recent paper by
Dordevic {\it et al.}\cite{dord2014} who attempted to factor out of the
superconducting state optical data, the direct effect of the superfluid
condensate and so access intrinsic quasiparticle properties. To accomplish
this they advocated subtracting first from the imaginary part of the
superconducting state optical conductivity, $\sigma^{sc}_2(T,\Omega)$
the Kramers-Kronig transform of the superfluid density $\rho_s$ which
appears in the real part of $\sigma^{sc}(T,\Omega)$ as a Dirac delta-function
at $\Omega=0$ of weight $\rho_s$. After this modification the optical
scattering rate which we denoted $1/\tau^{sc-m}_{\rm opt}(T,\Omega)$
was computed. First, at photon energies twice the maximum phonon energy plus
the optical gap energy (of the order of $26\,$meV for Pb)
we found that superconducting and BQP gas
scattering rates agree to within a few percent as we might
have expected. In contrast,
the sc-m case predicted that the scattering rate was larger by almost a
factor of two in this energy region. Second, when we compared with the
optical scattering rate $1/\tau^{n}_{\rm opt}(T,\Omega)$ obtained for
the normal state, we find good agreement with the
superconducting state and with $1/\tau^{bqp}_{\rm opt}(T,\Omega)$.
This demonstrated that in this energy range the superconducting optical
scattering rate, as has been widely defined in the literature, indeed
provided a reliable measure of relaxation in both normal and BQP gas.
This was not surprising since at these energies the superconducting
gap is small as compared with the total energy involved.

In the phonon assisted energy region between 10 and $20\,$meV
(in the specific case of Pb) a broad peak appeared in the optical
scattering rate of the superconducting state which was not seen in either
the normal state or the BQP gas. This translated into deviations of
$\sim20\,\%$, still rather modest. We also found that, at finite temperature,
a sharp peak developed in the optical scattering rate of the BQP gas below
the optical gap. Such a peak was neither present in the superconducting nor
in the normal state. It was traced to a zero crossing of the imaginary part
of the BQP gas' optical conductivity. It is a signature
of this crossing and is not directly related to a relaxation
rate.  Furthermore, from the optical gap energy to approximately twice
this energy the BQP optical scattering rate had a plateau which was traced
back to the interband transitions present in this case, even in the clean
limit. However, above $\sim 6\,$meV $[\sim4\Delta(T)]$ we found that
when $1/\tau^{bqp}_{\rm opt}(T,\Omega)$ was compared with
$1/\tau^{n}_{\rm opt}(T,\Omega)$ the agreement was
almost as good as it was at high energies. Thus, normal state
and BQP gas optical scattering rates were close to each other in the
entire range of photon energies above twice the optical gap.
This central conclusion remains true when finite residual scattering
due to disorder is introduced, when the temperature is increased,
when we go from strong coupling Pb with a soft phonon spectrum to
weak coupling V$_3$Si with a large maximum phonon energy of
$\sim 50\,$meV, and when gap anisotropy is considered. The specific
example provided is for Hg1201, a cuprate with $d$-wave gap
symmetry.

\appendix
\section{}
\label{app:A}

At zero temperature the real part of the dynamic optical conductivity
associated with the BQP gas is given by Eq.~\eqref{eq:1} with only the
$A({\bf k},\omega)A({\bf k},\omega+\Omega)$ term contained in the last
square bracket. We get
\begin{widetext}
\begin{equation}
 \label{eq:a1}
 \textrm{Re}\{\sigma^{bqp}(T=0,\Omega)\} = \frac{e^2}{2\Omega}v^2_FN(0)
 \int_{-\infty}^\infty\!d\epsilon\int_{-\Omega}^0\!d\omega\,
 A({\bf k},\omega)A({\bf k},\omega+\Omega).
\end{equation}
\end{widetext}
Here, $N(0)$ is the normal state electronic density of states
associated with energies $\epsilon\equiv\epsilon_{\bf k}$
taken at the chemical potential $\mu$ which we set to zero. In
the BCS clean limit, i.e. no residual scattering and zero inelastic
contribution, the spectral density $A({\bf k},\omega)$ reduces to
two Dirac delta-functions of the form:
\begin{equation}
 \label{eq:a2}
 A(\epsilon_{\bf k},\omega) = u^2_{\bf k}\delta(\omega-\epsilon_{\bf k})+
    v^2_{\bf k}\delta(\omega+\epsilon_{\bf k}),
\end{equation}
with
\begin{eqnarray}
   u^2_{\bf k } &=& \frac{1}{2}\left(1+\frac{\epsilon_{\bf k}}{E_{\bf k}}
   \right)\nonumber\\
   v^2_{\bf k } &=& \frac{1}{2}\left(1-\frac{\epsilon_{\bf k}}{E_{\bf k}}
   \right),
 \label{eq:a3}
\end{eqnarray}
where
\begin{equation}
 \label{eq:a4}
 E_{\bf k} = \sqrt{\epsilon^2_{\bf k}+\Delta^2}.
\end{equation}
Here $\Delta$ is the BCS gap and $2\Delta$ is the corresponding optical gap.

\begin{figure}[t]
\includegraphics[width = 95mm]{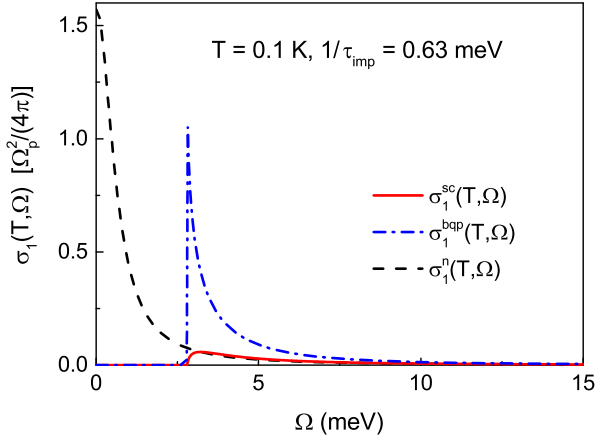}
\caption{(Color online) The BCS real part of the optical conductivity
$\sigma_1(T,\Omega)$ at $T=0.1\,$K and with elastic impurity scattering
$1/\tau_{\rm imp} = 0.63\,$meV. The superconducting gap $\Delta(0)=1.4\,$meV.
The solid (red) curve applies to the superconducting
state (sc), the dashed-dotted (blue) curve to the Bogoliubov quasiparticle gas
(bqp), and the dashed (black) curve to the normal state (n).}
\label{fig:fig8}
\end{figure}

Substituting Eq.~\eqref{eq:a2} for $A({\bf k},\omega)$ in Eq.~\eqref{eq:a1}
leads to an intraband and an interband contribution. The first is proportional
to $\delta(\Omega)$ while the second contribution has a $\delta(2\omega+\Omega)$
factor. Specifically, we get, setting $\Lambda\equiv e^2v_F^2N(0)/2$:
\begin{widetext}
\begin{eqnarray}
  \textrm{Re}\{\sigma^{bqp}_{\rm intra}(T=0,\Omega)\}&\propto&\frac{\Lambda}{\Omega}\delta(\Omega)
  \int_{-\infty}^\infty\!d\epsilon
  \int_{-\Omega}^0\!d\omega\,\left[u^4_{\bf k}\delta(\omega-E_{\bf k})+
    v^4_{\bf k}\delta(\omega+E_{\bf k})\right],\nonumber\\
  \textrm{Re}\{\sigma^{bqp}_{\rm inter}(T=0,\Omega)\}&\propto&\frac{\Lambda}{\Omega}
  \int_{-\infty}^\infty\!d\epsilon
  \int_{-\Omega}^0\!d\omega\,\delta(2\omega+\Omega)u^2_{\bf k}v^2_{\bf k}
  \left[\delta(\omega-E_{\bf k})+\delta(\omega+E_{\bf k})\right].
 \label{eq:a5}
\end{eqnarray}
\end{widetext}
The first term, $\textrm{Re}\{\sigma^{bqp}_{\rm intra}\}$ involves intraband
optical transitions and the second, $\textrm{Re}\{\sigma^{bqp}_{\rm inter}\}$,
interband transitions. In each term, the Dirac delta-function $\delta(\omega-E_{\bf k})$
never clicks in the range of integration for $\omega$ and we are left with
a single contribution involving only $\delta(\omega+E_{\bf k})$. The
interband contribution works out to be
\begin{equation}
  \textrm{Re}\{\sigma^{bqp}_{\rm inter}(T=0,\Omega)\} \propto 
  \frac{\Lambda}{\Omega^2}\frac{\Delta}{\sqrt{\Omega^2-(2\Delta)^2}},
 \label{eq:a6}
\end{equation}
for $\Omega > 2\Delta$ and zero for $\Omega < 2\Delta$. The intraband
contribution is zero as we expect at $T=0$. The interband transitions are
seen in the dashed-dotted (blue) curve of Fig.~\ref{fig:fig2}. On the
other hand, to get the real part of the dynamic optical conductivity
in the superconducting state we need to include the
$B({\bf k},\omega)B({\bf k},\omega+\Omega)$ term of Eq.~\eqref{eq:1}
and this leads to a complete cancellation of the interband
transitions, a well known effect for class II matrix elements.
As seen in the solid (red) curve of Fig.~\ref{fig:fig8}
$\sigma^{sc}_1(T,\Omega)$ does not show an inverse square root singularity
at $\Omega = 2\Delta$ but rather starts from zero at that point.
This behavior is different from that of the dashed-dotted
(blue) curve for the BQP gas and, of course, from the normal state
response [dashed (black) curve].

\section{}
\label{app:B}

It is most convenient to solve the Eliashberg equations as two coupled
non-linear equations in an imaginary axis representation:\cite{carb1990}
\begin{widetext}
\begin{subequations}
\label{eq:B1}
 \begin{equation}
  \tilde{\Delta}(T,i\omega_n) = \pi T \sum_m \left[\lambda(i\omega_n-
   i\omega_m)-\mu^\star\right]\frac{\tilde{\Delta}(T,i\omega_m)}
   {\sqrt{\tilde{\omega}^2(T,i\omega_m)+\tilde{\Delta}^2(T,i\omega_m)}},
 \end{equation}
and
 \begin{equation}
  \tilde{\omega}(T,i\omega_n) = \omega_n+\pi T\sum_m
  \lambda(i\omega_n-i\omega_m)\frac{\tilde{\omega}(T,i\omega_m)}
   {\sqrt{\tilde{\omega}^2(T,i\omega_m)+\tilde{\Delta}^2(T,i\omega_m)}}.
 \end{equation}
\end{subequations}
\end{widetext}
 Here, $T$ is the temperature, $\mu^\star$ is the Coulomb pseudo-potential,
 $\omega_n = \pi T(2n+1), n = 0,\pm 1,\pm 2,\ldots$ the Matsubara frequencies,
 and $\lambda(i\omega_n-i\omega_m)$ is given by:
 \begin{equation}
  \lambda(i\omega_n-i\omega_m) = 2\int_0^\infty\!d\Omega\,
  \frac{\Omega \alpha^2F(\Omega)}{\Omega^2+(\omega_n-\omega_m)^2}.
 \end{equation}
Eqs.~\eqref{eq:B1} are valid for any amount of electron-impurity scattering
as this contribution cancels in these equations. This set of equations can
be used to calculate thermodynamic properties, the penetration depth, etc.

If one wants to determine response functions, like the optical conductivity,
the renormalized frequencies $\tilde{\omega}(T,\omega)$ and the gap function
$\tilde{\Delta}(T,\omega)$ are required. These are obtained from Eqs.~\eqref{eq:B1}
by analytic continuation using an algorithm developed by F.~Marsiglio
{\it et al.}\cite{mars1988} and are denoted by $\tilde{\omega}_{\rm clean}(T,\omega)$
and $\tilde{\Delta}_{\rm clean}(T,\omega)$. If electron-impurity scattering described by
an impurity scattering rate $1/\tau_{\rm imp}$ is present, one has to
perform, in addition, the following renormalization:\cite{mars1997}
\begin{subequations}
\label{eq:B3}
 \begin{equation}
   \tilde{\omega}(T,\omega) = \tilde{\omega}_{\rm clean}(T,\omega)+\frac{i}{2\tau_{\rm imp}}
   \frac{\tilde{\omega}(T,\omega)}{\sqrt{\tilde{\omega}^2(T,\omega)-
   \tilde{\Delta}^2(\omega)}},
 \end{equation}
 and
 \begin{equation}
   \tilde{\Delta}(T,\omega) = \tilde{\Delta}_{\rm clean}(T,\omega)+\frac{i}{2\tau_{\rm imp}}
   \frac{\tilde{\Delta}(T,\omega)}{\sqrt{\tilde{\omega}^2(T,\omega)-
   \tilde{\Delta}^2(T,\omega)}},
 \end{equation}
\end{subequations}
which have to be solved self consistently.
The functions $\tilde{\omega}(T,\omega)$ and $\tilde{\Delta}(T,\omega)$
[either in the clean limit or renormalized according to Eqs.~\eqref{eq:B3}]
are then used in Eqs.~\eqref{eq:4} to \eqref{eq:6} to determine the optical
conductivity from Eq.~\eqref{eq:2}.

\acknowledgements
Research supported in part by the Natural Sciences and Engineering Research
Council of Canada (NSERC) and by the Canadian Institute for Advanced Research
(CIFAR). We thank E. J. Nicol for assistance with our Eliashberg programs.
%
% The bibliography
%
\bibliography{bqp}
\end{document}